\documentclass[sn-mathphys]{sn-jnl}% Math and Physical Sciences Reference Style
\jyear{2021}%

\theoremstyle{thmstyleone}%
\theoremstyle{thmstyletwo}%

\theoremstyle{thmstylethree}%

\usepackage{amsmath,amssymb}
\usepackage{hyperref}

\usepackage[normalem]{ulem}
\usepackage{xcolor}

\renewcommand{\v}[1]{\mathrm{\textbf{#1}}}
\newcommand{\av}[1]{\left\langle{#1}\right\rangle}
\newcommand{\avsmall}[1]{\langle{#1}\rangle}
\newcommand{\abs}[1]{\lvert{#1}\rvert}

\begin{document}

\title[Universal alignment in turbulent pair dispersion]{Universal alignment in turbulent pair dispersion}

%Universal alignment of turbulent pair dispersion

%Degrees of separation: Universal alignment of turbulent pair dispersion

%Degrees of separation: A persistent alignment of turbulent pair dispersion

%%=============================================================%%
%% Prefix	-> \pfx{Dr}
%% GivenName	-> \fnm{Joergen W.}
%% Particle	-> \spfx{van der} -> surname prefix
%% FamilyName	-> \sur{Ploeg}
%% Suffix	-> \sfx{IV}
%% NatureName	-> \tanm{Poet Laureate} -> Title after name
%% Degrees	-> \dgr{MSc, PhD}
%% \author*[1,2]{\pfx{Dr} \fnm{Joergen W.} \spfx{van der} \sur{Ploeg} \sfx{IV} \tanm{Poet Laureate} 
%%                 \dgr{MSc, PhD}}\email{iauthor@gmail.com}
%%=============================================================%%

\author*[1,2]{\fnm{Ron} \sur{Shnapp}}\email{ronshnapp@gmail.com}

\author[2,3]{\fnm{Stefano} \sur{Brizzolara}} \email{brizzolara@ifu.baug.ethz.ch}

\author[2,3]{\fnm{Marius M.} \sur{Neamtu-Halic}} \email{nemarius@ethz.ch}

\author[2,3]{\fnm{Alessandro} \sur{Gambino}}\email{agambino@ethz.ch}

\author[2,4]{\fnm{Markus} \sur{Holzner}}\email{holzner@ifu.baug.ethz.ch}

\affil[1]{\orgdiv{Department of Mechanical Engineering}, \orgname{Ben-Gurion University of the Negev}, \orgaddress{\city{Beer-Sheva}, \postcode{P.O.B. 653}, \country{Israel}}}

\affil[2]{\orgname{Swiss Federal Institute of Forest, Snow and Landscape Research WSL}, \orgaddress{\city{Birmensdorf}, \postcode{8903}, \country{Switzerland}}}

\affil[3]{\orgdiv{Institute of Environmental Engineering}, \orgname{ETH Z\"{u}rich}, \orgaddress{\city{Z\"urich}, \postcode{CH-8039}, \country{Switzerland}}}

%\affil[4]{\orgname{DIFI, University of Genova and INFN, Genova Section}, \orgaddress{\street{Via Dodecaneso 33}, \city{Genova}, \postcode{I-16146}, \country{Italy}}}

\affil[4]{\orgname{Swiss Federal Institute of Aquatic Science and Technology Eawag}, \orgaddress{\city{D\"ubendorf}, \postcode{8600}, \country{Switzerland}}}

\abstract{Countless processes in nature and industry, from rain droplet nucleation to plankton interaction in the ocean, are intimately related to turbulent fluctuations of local concentrations of advected matter. These fluctuations can be described by considering the change of the separation between particle pairs, known as pair dispersion, which is believed to obey a cubic in time growth according to Richardson's theory. Our work reveals a universal, scale-invariant alignment between the relative velocity and position vectors of dispersing particles at a mean angle that we show to be a universal constant of turbulence. We connect the value of this mean angle to Richardson's traditional theory and find agreement with data from a numerical simulation and a laboratory experiment. While the Richardson's cubic regime has been observed for small initial particle separations only, the constancy of the mean angle manifests throughout the entire inertial range of turbulence. Thus, our work reveals the universal nature of turbulent pair dispersion through a geometrical paradigm whose validity goes beyond the classical theory, and provides a novel framework for understanding and modeling transport and mixing processes.}

%\keywords{keyword1, Keyword2, Keyword3, Keyword4}

%%\pacs[JEL Classification]{D8, H51}

%%\pacs[MSC Classification]{35A01, 65L10, 65L12, 65L20, 65L70}

\maketitle

\section{Introduction}

If we took a handful of small passive particles and threw them into the ocean, how long would it take before the particles became fully mixed in the oceans across the globe? Answering questions like this requires that we know the rates at which turbulent flows transport and diffuse the materials that they carry. One characteristic of transport is the so-called pair dispersion, which describes the rate at which two particles separate from each other. Pair dispersion can be used to calculate the variance of the concentration fluctuations of substances carried by the flow~\cite{Batchelor1952}, so it is of critical importance in numerous applications such as determining the rate of ozone destruction in the atmosphere~\cite{Edouard1996} or the dispersion of pollutants in the ocean~\cite{Griffa2013}.

\begin{figure}[b]
	\centering
	\includegraphics[width=90mm]{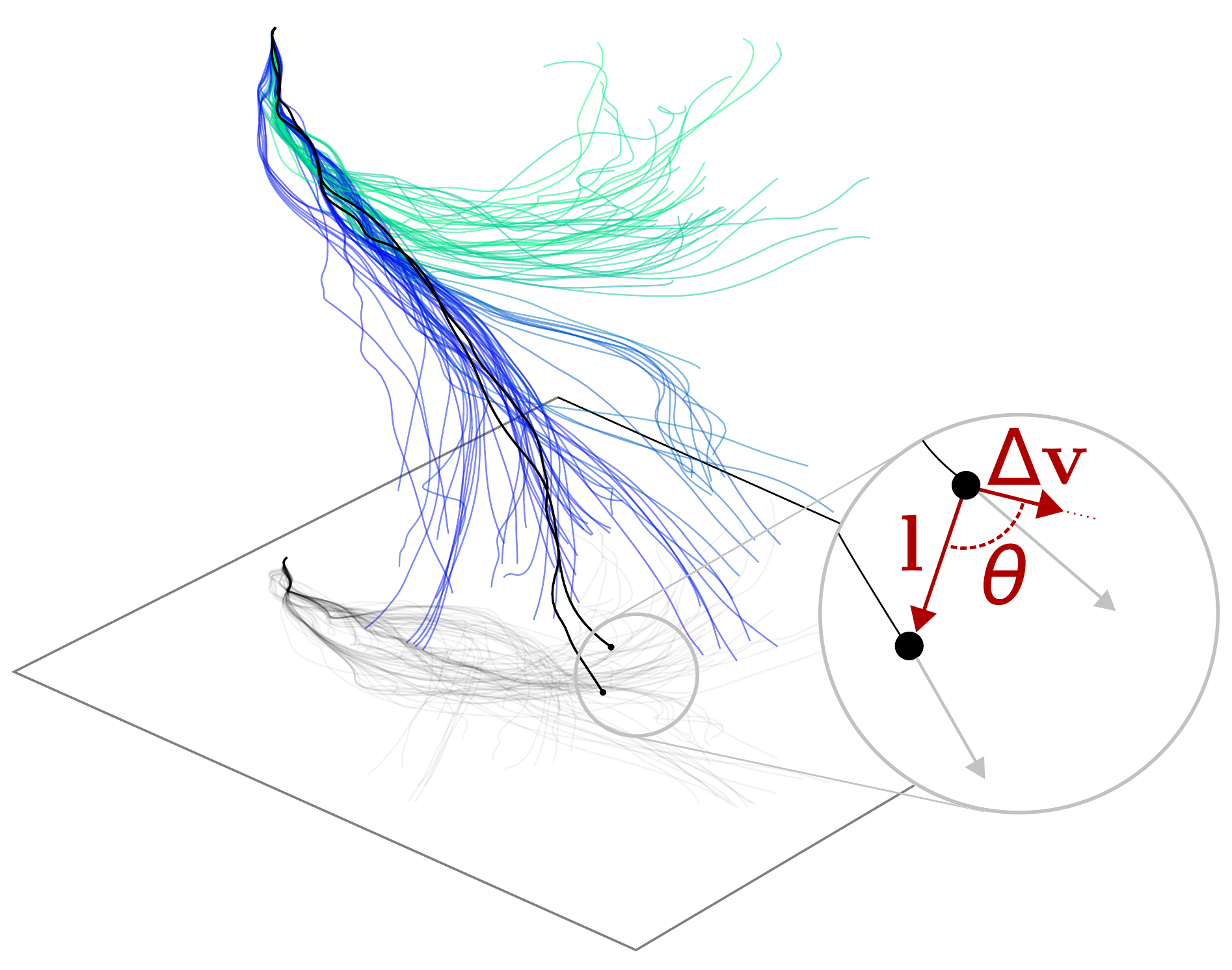}
	\caption{\textbf{Pair dispersion angle definition.} Visualization of a group of Lagrangian fluid particle trajectories in a turbulent flow, starting from within a cube with dimensions of one Kolmogorov length scale. Trajectories are color-coded according to their final position. The relative position vector, the relative velocity, and the angle between them, $\v{l}$, $\Delta \v{v}$, and $\theta$, respectively, are shown for one particle pair. In the inertial range, the average of $\theta$ calculated over all pairs is equal to a constant value. \label{fig:def}}
\end{figure}

In quiescent fluids particles undergo Brownian motion, so the increase of the distance between two particles is a diffusive process with constant diffusivity. However, in turbulence, particles are transported by the flow field and so the process is driven by advection. Turbulent pair dispersion is divided into three regimes (see reviews in Refs.~\cite{Monin1972, Sawford2001, Salazar2009}). First, at very short times, the finite fluid inertia dictates that the relative velocity of particles remains approximately constant. This leads to the so-called "ballistic" regime, in which the inter-particle distance scales linearly with time~\cite{Batchelor1952}. The ballistic regime occurs at times $t<\tau_b$, where $\tau_b=(l_0^2/\epsilon)^{1/3}$ is the Batchelor time scale ($l_0$ is the initial distance between particles and $\epsilon$ is the mean dissipation rate of kinetic energy into heat). The second, so-called "diffusive" regime of separation ensues at very long times when the inter-particle distances, $l$, are outside the inertial range, $l \gg L$ (where $L$ is the integral length scale). There, similarly to the Brownian motion case, the velocity field is uncorrelated and the typical inter-particle distance grows with a square-root scaling in time~\cite{Taylor1921}. The third regime was introduced by Richardson nearly a century ago~\cite{Richardson1926}, and it is still intensely debated today. This regime corresponds to the inertial range of turbulence, namely, $\eta \ll l \ll L$, where $\eta=(\nu^3/\epsilon)^{1/4}$ is the Kolmogorov length scale ($\nu$ is the kinematic viscosity). In this inertial regime the typical separation velocities increase with the separation distance, which leads to a super-diffusive growth of inter-particle distances.

While the validity of the ballistic and the diffusive regimes are generally agreed upon by the community, Richardson's inertial regime remains contested.   
Indeed, although Richardson's theory was proposed nearly a century ago~\cite{Richardson1926}, and despite it being considered a hallmark of turbulence, it remains elusive to empirical verification.
In particular, finite Reynolds number effects~\cite{Yeung1994, Ott2000, Biferale2005, Bourgoin2006, Ouellette2006, Buaria2015, Shnapp2018, Tan2022, Elsinga2022}, and intermittency and mixing of different regimes~\cite{Boffetta2002, Yeung2004, Scatamacchia2012}, make it difficult to interpret the measurements of the time evolution of the statistics of inter-particle distances over long periods of time.
Indeed, empirical studies that measured pair dispersion scaling exponents found agreement with Richardson's theory only for particular choices of initial conditions, namely small initial distances~\cite{Ott2000, Berg2006, Elsinga2022, Tan2022, Kishi2022} or separation velocities~\cite{Shnapp2018}, while for other initial conditions the scaling exponent were different. 
This gap between the classical theory and modern experiments demands an explanation that will better describe the process across the scales.

In this paper, we propose a new approach for characterizing turbulent pair dispersion. Specifically, we focus on a new observable: the angle, $\theta$, formed between the relative position and relative velocity vectors (see Fig.~\ref{fig:def}, and Supplementary Video 1). We show theoretically and confirm empirically that this angle has three unique behaviors in each of the three regimes of the separation process. In particular, in the inertial regime it is constant and equal to approximately 59$^\circ$ independently of the initial conditions and the Reynolds number. Thus, it is a universal constant of turbulence. Furthermore, we calculate the mean value of the angle analytically using Richardson's theory at small initial separations, finding agreement with the empirical data. 
Thus, our work introduces a geometrical framework that reveals the universality of turbulent pair dispersion and applies more broadly than the traditional picture. This provides a new framework for characterizing dispersion in oceanic and atmospheric turbulent flows.

\section{Results}

\subsection{Theoretical analysis of the pair dispersion angle}\label{sec:theory}

We begin by calculating theoretical predictions for the average of $\theta$ in turbulence.
To elucidate our analysis, we will consider three different scenarios: first, particles that move with constant velocities, second, particles that undergo normal diffusion, and third, particles that undergo a super-diffusive separation. These three scenarios correspond to the three regimes of pair dispersion in turbulence namely, the ballistic, diffusive and inertial regimes respectively. 
We consider particles that have initial positions $\v{x}_1(0)$ and $\v{x}_2(0)$, and move with velocities, $\v{v}_1(t)$ and $\v{v}_2(t)$, respectively. Their relative velocity is $\Delta \v{v}(t) = \v{v}_1(t)-\v{v}_2(t)$. The distance between them, $l(t) \equiv \abs{\v{l} (t)} = \abs{\v{x}_1(t) - \v{x}_2(t)}$, follows the kinematic relation~\cite{Yeung1994, Tsinober2009} 
	\begin{equation}
	\frac{d\, l}{dt} = \frac{\Delta \v{v} \cdot \v{l}}{l} \,\, .
	\label{eq:dldt}
	\end{equation}
The cosine of the angle that lies between the relative velocity and relative position vectors is equal to
	\begin{equation}
	\cos(\theta) \equiv \alpha  = \frac{\Delta \v{v} \cdot \v{l}}{\abs{\Delta\v{v}}  \, l} \,\, ,
	\label{eq:alpha}
	\end{equation}
where it can take values between -1 and 1, which correspond to $\theta$ ranging from $180^\circ$ to $0^\circ$. Combining eq.~\eqref{eq:dldt} and \eqref{eq:alpha} shows that
\begin{equation}
\alpha = \frac{1}{\abs{\Delta \v v}} \frac{d\,l}{dt} \\ .
\label{eq:alpha_2}
\end{equation}

In the first scenario, the ballistic case, the particles' velocities are frozen in time. Thus, the relative position in this regime evolves as $\v{l}(t) = \v{l}(0) + \Delta \v{v}\,t$, and the angle cosine can be solved analytically, giving
$$\alpha(t)=  \frac{\Delta \v{v} \cdot (\v{l}(0)+\Delta\v{v}\,t)}{ \abs{\v{l}(0) + \Delta\v{v} \, t} \, \abs{\Delta \v{v}}} \,\, .$$
Therefore, in the ballistic scenario $\alpha(t)$ increases monotonically and tends asymptotically towards 1 for any $\v l(0)$ and $\Delta \v v$. Correspondingly, the two vectors tend towards perfect alignment with time, namely, $\theta \rightarrow 0^{\circ}$.

We move on to the second scenario by analyzing the time evolution of $\av{\alpha}$ (where brackets denote an ensemble average) for an ensemble of particles that undergo diffusion with a constant diffusivity. Specifically, we consider an ensemble composed by taking pairs of particles at different locations and times with the same initial separation $\v{l}(0)=\v{l}_0$, while allowing them to separate with time. Initially, there is no preferred alignment ($\av{\alpha(0)}=0$) since the particles' velocities are chosen randomly. Furthermore, the particles separate from each other on average, so $\av{\frac{dl}{dt}}>0$; together with eq.~\eqref{eq:alpha_2}, this suggests that $\av{\alpha(t)}$ immediately grows and becomes positive for $t>0$. To determine the long time behavior we transform to a frame of reference whose origin is fixed at $\v{x}_1(t)$. Then, considering long times for which $l(t) \gg l_0$ (where $l_0\equiv \abs{\v{l}(0)}$), the average distance between pairs in three dimensional space is given by $\avsmall{l(t)} = (\frac{16}{\pi}\,D\,t)^{1/2}$, where $D$ is the diffusivity for the relative dispersion, equal to twice the single particle diffusivity. Taking the time derivative and using eq.~\eqref{eq:dldt} and \eqref{eq:alpha}, we obtain the long-time scaling 
$\av{\alpha(t)} \sim \sqrt{\frac{4\,D}{\pi \av{\abs{\Delta \v{v}}}^2}} \,t^{-1/2}$.
Thus, at long times, $\av{\alpha}$ decays asymptotically back towards zero with a time scaling of $t^{-1/2}$, so that $\av{\theta(t)}\rightarrow 90^{\circ}$. The asymptotic decay in this case occurs because $\avsmall{l}$ grows with time with a scaling exponent smaller than 1, while the statistics of $\Delta \v{v}$ are constant.

Moving forward to the third scenario, pair dispersion in the inertial range of turbulence is intrinsically different from the above cases, since both the scaling of $\avsmall{l}$ is different, and statistics of $\Delta \v{v}$ depend on the scale $l$. Indeed, a preferential oblique alignment between $\v{l}$ and $\Delta \v{v}$ was observed in direct numerical simulations (DNS)~\cite{Yeung1994, Yeung2004}. To calculate the mean value of $\alpha$ we use eq.~\eqref{eq:alpha_2} and expand it as a Taylor series around the mean values of $\abs{\Delta v}$ and $\frac{d\,l}{dt}$. By averaging the series, we obtain the following expression for the mean value
\begin{equation}
\av{\alpha}=\av{\frac{1}{\abs{\Delta \v v}} \frac{d\,l}{dt}} = \alpha_0 \left(1 + \sum\limits_{n=1}^{\infty} \frac{\alpha_n}{\alpha_0}\right)
\label{eq:A}
\end{equation}
where $\alpha_n$ is the averaged n$^{\mathrm{th}}$ term of the series, and $\alpha_0 = \frac{1}{\av{\abs{\Delta \v v}}} \frac{d\,\av{l}}{dt}$ (the derivation is given in Sec.~\ref{sec:alpha_expansion}). In what follows, we assume that the series converges sufficiently fast and truncate terms with $n\geq2$. This results in a first order approximation of eq.~\eqref{eq:A}, the accuracy of which is confirmed in Sec.~\ref{sec:results}. Then, since $\alpha_1 = 0$, we obtain that $\av{\alpha}=\alpha_0$.

The behavior of $\alpha_0$ could first be considered from a dimensional analysis point of view. In the inertial range, Kolmogorov's local isotropy theory leads to the conclusion that $\alpha_0$ can only depend on $\epsilon$, $l_0$, and $t$. In addition to that, $\alpha$ does not depend on $l$ explicitly but only on $\frac{dl}{dt}$ (eq.~\eqref{eq:alpha_2}). Therefore, and because $\frac{d\av{l}}{dt}=\frac{d\av{l-l_0}}{dt}$, it is reasonable to assume that $\alpha_0$ is independent on $l_0$ in the inertial range. This leaves only $\epsilon$ and $t$ as the parameters of the problem in the inertial range. However, no dimensionless group can be constructed from these two parameters, so to ensure dimensional homogeneity, $\av{\alpha}$ has to be constant. The actual validity of this assumption is verified below (Sec.~\ref{sec:results}).

We can also calculate $\alpha_0$ using Richardson's theory in the range in which it is valid. Thus, we calculate $d\av{l}/dt$ and $\av{\abs{\Delta \v v}}$ in the inertial range. When $l(t)$ is in the inertial range and sufficient time has passed such that $\avsmall{l(t)} \gg l_0$, Richardson's law predicts the following super-diffusive pair dispersion regime~\cite{Richardson1926, Salazar2009}
\begin{equation}
\avsmall{l^2} = g \, \epsilon \, t^3 \,\, ,
\label{eq:Richardson}
\end{equation}
where $g$ is the Richardson constant. We also make use of the fact that there are two theoretical predictions for the probability density function (PDF) of $l$ in the inertial range~\cite{Richardson1926, Batchelor1952, Salazar2009, Ouellette2006}. In both theories, the average separation is equal to $\avsmall{l}=b\,\sqrt{\avsmall{l^2}}$, where $b=0.867\pm0.054$ is a dimensionless constant (see Sec.~\ref{sec:calculating_b}). Combining this with eq.~\eqref{eq:Richardson} we obtain $\frac{d\av{l}}{dt} = \frac{3}{2}\,b\,(g\,\epsilon\,t)^{1/2}$. Notably, Richardson's solution is strictly valid only for $l_0=0$, and thus the variance of $l$ in eq.~\eqref{eq:Richardson} is usually replaced with the variance of $(l-l_0)$~\cite{Salazar2009}. Yet, since $\frac{d\av{l}}{dt}=\frac{d\av{l-l_0}}{dt}$, and since the PDF of $l-l_0$ was experimentally observed to agree with the theoretical expressions used here for the PDF of $l$ \cite{Ouellette2006, Salazar2009}, our calculation applies also for finite $l_0$.	
The second factor, $\av{\abs{\Delta \v{v}}}$, is the first order Eulerian-Lagrangian absolute structure function, where the relative velocities are taken over the full distribution of particle distances which changes with time. At $t=0$, the structure function is purely Eulerian, so according to the Kolmogorov theory~\cite{Kolmogorov1941} (namely, neglecting intermittency corrections), $\av{\abs{\Delta \v{v}} \, \big\lvert\, t=0} = \av{ \abs{\Delta_{l_0} \v{v}}} = C_1 (\epsilon\,l_0)^{1/3}$, where $C_1$ is a universal constant of turbulence. At longer times, the mixed structure function is calculated by averaging the particles' relative velocities across the distribution of particle distances, $l$; in the inertial range we obtain $\av{\abs{\Delta \v{v}}} = c \, C_1 \big(\epsilon \, \sqrt{\av{l^2}} \big)^{1/3}$, where $c = 0.918\pm0.034$ is a dimensionless constant (see Sec.~\ref{sec:Eu-Lag_SF}). Combining these estimations and using eq.~\eqref{eq:Richardson} we obtain the following first order, mean-field approximation, for the angle cosine in the inertial range of turbulence 
\begin{equation}
\av{ \alpha } = 
a\, \frac{g^{1/3}}{C_1} \,\, ,
\label{eq:alpha_Richardson}
\end{equation}
where $a \equiv \frac{3b}{2c} =  1.42 \pm 0.10$.

Equation~\eqref{eq:alpha_Richardson} connects $\av{\alpha}$ with Richardson's law in the inertial range, and it has several important implications. 
First, our calculations suggest that $\av{\alpha}$ in the super-diffusive regime is constant. This is in agreement with the dimensional analysis argument presented above, which is expected as eq.~\eqref{eq:Richardson} can also be derived from a similar argument~\cite{Monin1972}. Indeed, the value of $\av{\alpha}$ obtained here does not depend on $\epsilon$ nor on the initial conditions, so it is a universal constant of turbulence. The value of $\av{\alpha}$ can be calculated with eq.~\eqref{eq:alpha_Richardson} for the small initial separations for which Richardson's theory holds, and then, assuming that $\av{\alpha}$ is independent on $l_0$, the same value should hold for the entire inertial range (this is verified in Sec.~\ref{sec:results}).
Second, because of the geometrical constraint $\abs{\alpha} \leq 1$, and since all the constants in eq.~\eqref{eq:alpha_Richardson} are positive, we obtain a constraint for the value of the Richardson constant
\begin{equation}
g \leq \left(\frac{C_1}{a}\right)^{3} \,\, .
\label{eq:constraint}
\end{equation}
Third, if one measures $\av{\theta}$ from empirical data, the value of $g$ can be readily calculated.

\subsection{Universality of the pair dispersion angle} \label{sec:results}

\begin{figure}
	\centering
	\includegraphics[width=.95\linewidth]{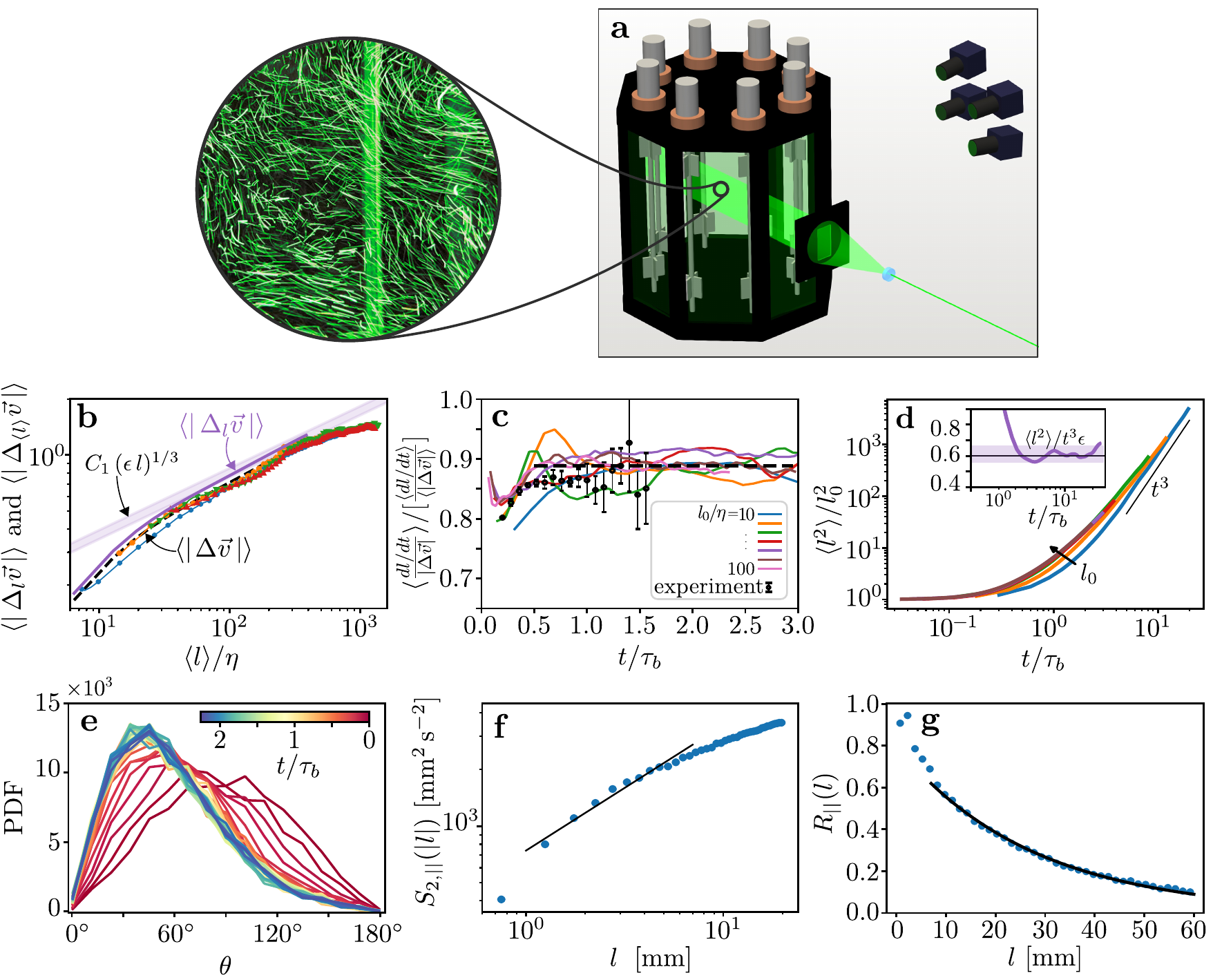}
	\caption{\textbf{Turbulence characterization.} \textbf{a} A sketch showing the experimental water tank, four camera system, laser light, and a streak image of flow tracers in the apparatus.
		\textbf{b} The Eulerian first-order absolute structure function is shown as a purple line, and a shaded region shows the Kolmogorov scaling with $C_1=2.7 \pm 0.15$. The absolute, first order, mixed Eulerian-Lagrangian structure function is shown as shapes for particle groups with $l_0/\eta=5,\,15,\,25, \,35 \pm 5$, and the black dashed line shows our theoretical prediction (eq.~\eqref{eq:mixed_Eu_Lag_structurefunction}). Data shown was taken from the DNS. 
		\textbf{c} The correction factor due to the series truncation in eq.~\eqref{eq:A}, plotted against time for data taken from the DNS for $l_0/\eta \in (10,20,30,40,50,70,100)$, and the experimental results, averaged over $l_0$ in the inertial range with uncertainty calculated by bootstrapping over three groups.
		\textbf{d} Mean squared particle separation is plotted against time for $\l_0/\eta<5$, $5<\l_0/\eta<10$, $10<\l_0/\eta<20$, $40<\l_0/\eta<50$, $60<\l_0/\eta<70$, and $90<\l_0/\eta<100$. The inset shows the same property for particles with $2<l_0/\eta<4$, normalized according to eq.~\eqref{eq:Richardson}, where $g=0.52\pm0.06$ is shown as a shaded region. Data shown uses the DNS results. 
		\textbf{e} The PDF of $\theta$ is shown at various times as indicated by the color of the curves. Data is shown for $l_0=50\eta$ using the DNS results. 
		\textbf{f} The Eulerian second-order longitudinal structure function in the experimental dataset is plotted against the distance, $l$. The black line and shaded region show the Kolmogorov scaling range for $10\eta < l < 40\eta$.
		\textbf{g} The longitudinal spatial correlation function of velocity fluctuations, averaged over spherical shells of radius $l$ in the experimental data set. The line shows an exponential fit to the tail of the data.
		\label{fig:3}}
\end{figure}

The angle $\theta$ can be measured from the trajectories of flow tracers, and thus, its behavior can be tested using empirical data directly. To this end, we used two independent datasets. The first is the Johns Hopkins Turbulence Database (JHTDB), which holds turbulent flow fields taken from a direct numerical simulation (DNS) of a forced homogeneous isotropic turbulence at a Taylor microscale Reynolds number of $\text{Re}_\lambda\approx 433$, with the ability to integrate Lagrangian trajectories~\cite{Li2008, Yu2012}. Since its publication, this database has become a gold standard and a hypothesis-testing tool for turbulent flows. The second data set was taken from 3D particle tracking measurements~\cite{Dracos1996, Shnapp2022} of quasi-homogeneous isotropic turbulence that we conducted inside a stirred water tank at ETH Z\"{u}rich (Fig.~\ref{fig:3}a, the data is available in Ref.~\cite{Shnapp2022a}). The flow had secondary circulation with an amplitude of about 68\% of the root mean squared turbulent fluctuations. The turbulence integral length scale, $L=20.5$ mm, was calculated by fitting an exponential function to the longitudinal velocity autocorrelation function (Fig.~\ref{fig:3}g), where a Kolmogorov scaling range was observed between approximately 1 and 5 mm for the Eulerian second order structure function (Fig.~\ref{fig:3}f). The Reynolds number was $\text{Re}_\lambda \approx 188$. Detailed information about both data sets are given in Sec.~\ref{sec:methods}.

We begin by evaluating the error that results from truncating the Taylor series in eq.~\eqref{eq:A}. The ratio between $\av{\alpha}$ and its first order approximation varies slightly with time where the same trend is observed for both data sets. At $t=0$ it equals approximately 0.8, and it then increases with time, plateauing at approximately 0.89  for $t\gtrsim\tau_b$ (Fig.~\ref{fig:3}c). Therefore, the error introduced by truncating the series amounts to approximately 10\% of the value of $\av{\alpha}$ and it does not change with time. Since the correction is constant throughout the inertial range, truncating the series does not affect the constancy predicted for $\av{\theta}$. 
This observation might be explained by the fact that the higher order terms result from correlations between $l$ and $\Delta\v v$ (Sec.~\ref{sec:alpha_expansion}), which should not change with time in the self-similar inertial range.

Next we confirm our calculation for the first order absolute structure function. The Eulerian structure function, shown in Fig.~\ref{fig:3}b, is compared with the Kolmogorov scaling, where neglecting intermittency corrections to the scaling exponent we obtain $C_1=2.7\pm0.15$. Furthermore, plotting the mixed Eulerian-Lagrangian structure function we obtain good agreement with eq.~\eqref{eq:mixed_Eu_Lag_structurefunction}. 
Overall, Fig.~\ref{fig:3}b and c confirm the hypotheses made in the derivation of eq.~\eqref{eq:alpha_Richardson}, and suggest that the error due to the Taylor series truncation is reasonably small.

We now turn to estimating the Richardson constant, $g$. The growth of $\avsmall{l^2}$ as a function of time is shown in Fig.~\ref{fig:3}d for particles taken from the DNS results. As commonly observed, we can identify a range in which eq.~\eqref{eq:Richardson} holds only for pairs with small initial separations~\cite{Ott2000, Elsinga2022, Tan2022}. Therefore, using particles with the initial separation $2\eta<l_0<4\eta$, we estimate $g=0.52\pm0.06$ (Fig.~\ref{fig:3}d, inset), which is in good agreement with previous measurements~\cite{Ott2000, Salazar2009}.
Furthermore, using the value we measured for $C_1$, eq.~\eqref{eq:constraint} gives $g\leq6.91$, which agrees with our measurements and with previous estimates~\cite{Salazar2009}.

\begin{figure}
	\centering
	\includegraphics[width=.95\linewidth]{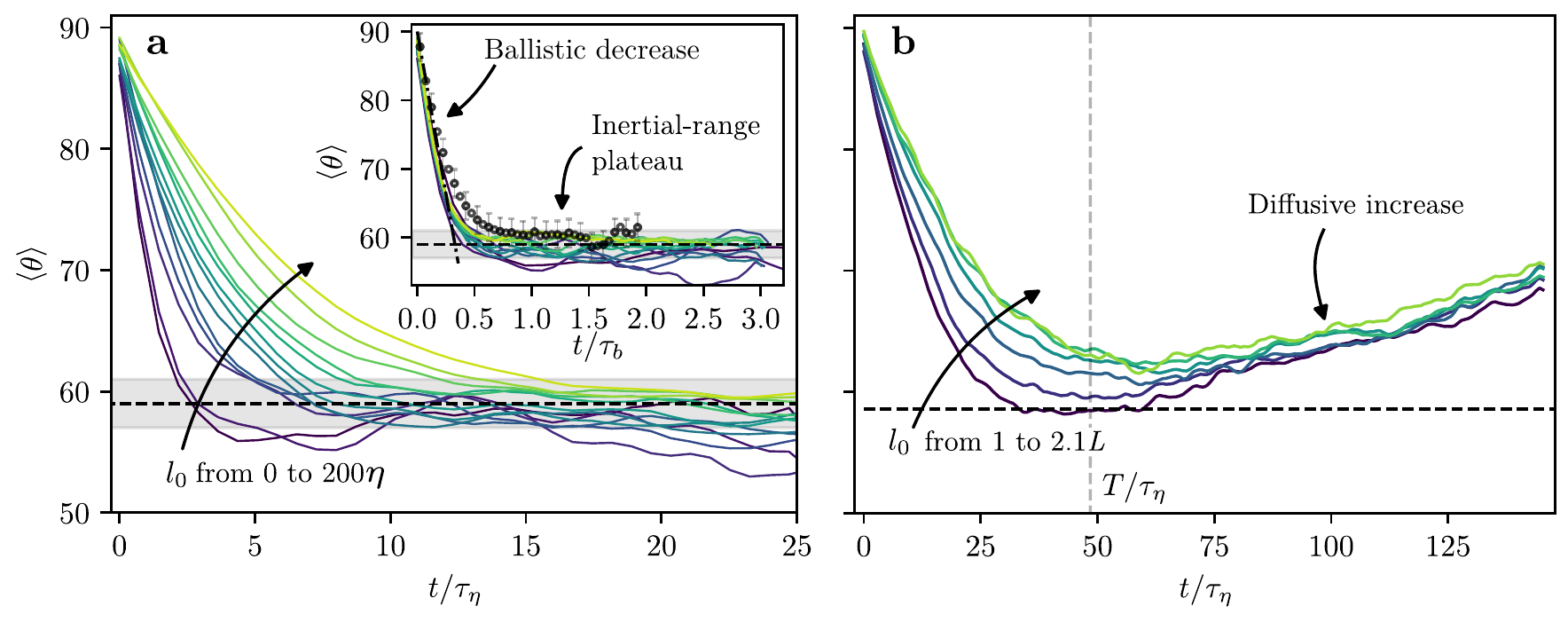}
	\caption{\textbf{Evolution of the pair dispersion angle.} \textbf{a}, Evolution of the average angle between the separation and the relative velocity vectors as a function of time. Continuous lines show DNS results for various ensembles grouped by the initial separation distance for values in the inertial range; the bin edges used to form the ensembles are $l_0/\eta = 0$, 10, 20, 30, 40, 50, 60, 70, 80, 90, 100, 130, 160, and 200, where the arrow runs from lower to higher values. Lines in the inset show the same data plotted with time normalized by the Batchelor timescale. Circles correspond to the experimental results, averaged over all pairs with $r_0<70\eta$ with an uncertainty of $\pm2^\circ$ based on the data range across the $l_0$ groups (Fig. S3). The dot-dashed line correspond to eq.~\eqref{eq:theta_initial}. The dashed black line and the shaded region mark the estimated value of $\av{\theta}=59.3 \pm 2^\circ$. 
	\textbf{b}, Evolution of the average of $\theta$ for pairs with initial separation outside the inertial range. Data are shown for initial separation distances of $l_0/\eta = 483$, 594, 704, 812, 915, and 1006. These values correspond to $l_0/L = 1.00,$ 1.24, 1.47, 1.69, 1.91, and 2.10. The horizontal dashed line marks the $\av{\theta}=59.3^\circ$ value, and the vertical dashed line marks one integral timescale. \label{fig:theta_vs_time}
		\label{fig:theta_vs_time_largescales}}
\end{figure}

Next we turn to measure the pair dispersion angle for particles in the empirical data sets using eq.~\eqref{eq:alpha} directly. The main panel of Fig.~\ref{fig:theta_vs_time}a shows the evolution of $\av{\theta}$ as a function of time for twelve pair ensembles from the DNS dataset, whose initial separation distance is inside the inertial range ($\eta<l_0<L$). For all cases, the average angle is initially very close to $90^\circ$ (values slightly smaller than $90^\circ$ are due to the well-known skewness of Eulerian velocity differences in turbulence~\cite{Monin1972}). Correspondingly, the PDF of $\theta$, that ranges from $0^\circ$ to $180^\circ$, is symmetrical around $90^\circ$ (Fig.~\ref{fig:3}e). Indeed, since the particles are chosen randomly, there is no significant preferred alignment between $\v{l}$ and $\Delta \v{v}$ at $t=0$.

For $t>0$, the average angle rapidly decreases until a plateau is reached where the time of convergence of $\av{\theta}$ to the plateau increases with the initial separation distance. In the ballistic regime, $t\ll\tau_b$, during which $\Delta \v{v}$ is constant, simple geometrical considerations show that
$\theta=\frac{\pi}{2}-\frac{\abs{\Delta_{\perp} \v{v}}}{\abs{l_0}}t$, 
where $\Delta_{\perp} \v{v}$ is the transverse component of the relative velocity. Taking the average of this relation over particles with the same $l_0$ while using the Kolmogorov scaling of the velocity differences we obtain the solution for short times
\begin{equation}
\av{\theta} = \frac{\pi}{2}\left( 1 -  \frac{t}{\tau_b} \, \frac{2\,C_{1,\perp}}{\pi}\right) \quad \text{for} \quad t \ll \tau_b \,\, ,
\label{eq:theta_initial}
\end{equation}
where $C_{1,\perp}\approx1.66$ is the coefficient for the transverse absolute first-order structure function ($\avsmall{\abs{\Delta_\perp v}}$) measured from our data. Therefore, to compensate for the differences in the convergence rate for pairs with different $l_0/\eta$, time is normalized by $\tau_b$ for the DNS and the experimental datasets. Following this normalization, the data from the DNS and from our experiment collapse, and good agreement is found with eq.~\eqref{eq:theta_initial} (Fig.~\ref{fig:theta_vs_time}a inset). The plateau is reached at $t/\tau_b \gtrsim 0.5$, approximately the time that marks the beginning of the super-diffusive regime~\cite{Bitane2012}. Since in ballistic motion $\av{\theta}$ decreases monotonically, and since the transition in Fig.~\ref{fig:theta_vs_time}a occurs at fixed $t/\tau_b$, we infer that the transition in the trends of $\av{\theta}$ is due to the transition of pair dispersion from the ballistic to the inertial, super-diffusive regime. Therefore, the plateau observed for $\av{\theta}$ in Fig.~\ref{fig:theta_vs_time}a occurs during the turbulent inertial regime; this confirms our prediction that $\av{\theta}$ attains a constant value in the inertial range, and that a plateau marks the inertial regime. Correspondingly, the PDF of $\theta$ becomes increasingly asymmetrical through the ballistic regime since its mode shifts to the left; eventually it reaches a steady-state for $t/\tau_b \gtrsim 0.5$, namely in the inertial range (Fig.~\ref{fig:3}e).

The value of the plateau of $\av{\theta}$ does not show a systematic dependence on $l_0$ nor on $\epsilon$, which is in accordance with the assumption presented in Sec.~\ref{sec:theory} and with eq.~\eqref{eq:alpha_Richardson}. In particular, the difference between the plateaus observed in the DNS and the experimental data is smaller than 2$^{\circ}$ ($58.6^\circ$ in the DNS and $60.3^\circ$ in the experiment), which is comparable to our uncertainty levels. This observation supports our prediction that the value of $\av{\theta}$ (and $\av{\alpha}$) in the inertial range is a universal constant of turbulence. Considering times $t>\tau_b$, we calculated the average value of the angle across $l_0$ averaging over both datasets, which gives an estimate of $\av{\theta}=59.3^\circ$ with an estimated uncertainty of $\pm2^\circ$ (Fig.~\ref{fig:theta_vs_time}a).

Using the values that we measured for $C_1$ and $g$, and plugging them into eq.~\eqref{eq:alpha_Richardson}, we can predict the value of the average angle cosine $\av{\alpha}_{th} = 0.422 \pm 0.042$ (the uncertainty reflect uncertainty in the values of $C_1$ and $g$). Furthermore, using the direct empirical measurement, we calculate that $\av{\alpha}_{meas} = 0.46 \pm 0.03$ (here the uncertainty reflects the slight variance in the experimental and numerical estimates).
Evidently, these two independent measurements are in good agreement given the 10\% accuracy of the first order approximation we used. This confirms eq.~\eqref{eq:alpha_Richardson} for the available degree of uncertainty.

In contrast to the behavior in the inertial range, Fig.~\ref{fig:theta_vs_time_largescales}b shows the evolution of $\av{\theta}$ for pairs outside the inertial range, with $l_0>L$ using the DNS data. Particles with such a large value of the initial separation are in the diffusive, Taylor range of turbulent pair dispersion, since velocity fluctuations are no longer correlated at these scales~\cite{Monin1972}. As for the inertial range case (Fig.~\ref{fig:theta_vs_time}a), $\av{\theta}$ is initially close to 90$^\circ$ and decreases rapidly. However, unlike in the inertial range case, here the curves do not plateau; instead, they reach local minima with values that depend on $l_0$, and then increase with time. This behavior of $\av{\theta}$ agrees with the prediction for the trend of $\av{\theta}\rightarrow90^\circ$ for diffusing particles (second scenario in Sec. 2). Overall, our observations confirm that in turbulence, only pairs in the super-diffusive pair dispersion regime manifest the plateau of average pair dispersion angle.

\section{Discussion}

Our work offers a framework which reveals the universality of turbulent pair dispersion. We show that the angle between the separation and the relative velocity vectors of separating particles is biased towards oblique values, and the evolution of its average, $\av{\theta}$, follows three distinct regimes in homogeneous isotropic turbulence. Starting from an unbiased value of approximately $90^{\circ}$, $\av{\theta}$ decreases linearly during the initial ballistic regime, while in the diffusive regime $\av{\theta}$ increases back towards the unbiased value of $90^\circ$. Yet, the key discovery of our work is that in the inertial range of turbulence $\av{\theta}$ plateaus at a constant value of approximately $59^\circ$ independently of the initial conditions and of the dissipation rate, making $\av{\theta}$ a universal constant of turbulence.

In Sec.~\ref{sec:theory}, we employed Richardson's classical theory~\cite{Richardson1926} to estimate the mean value of $\alpha$ (namely, $\cos\,\theta$) analytically, arriving at eq.~\eqref{eq:alpha_Richardson}. Nevertheless, the validity of Richardson's law with the unique $t^3$ scaling is observed only for small initial separations or separation velocities~\cite{Ott2000, Berg2006, Shnapp2018, Elsinga2022, Tan2022, Kishi2022}, while our work establishes that the constancy of $\av{\theta}$ holds throughout the entire inertial range (Fig.~\ref{fig:theta_vs_time}). To confirm the validity of the estimate provided by eq.~\eqref{eq:alpha_Richardson}, we measured the Richardson constant, $g$, using particles with initial separations in the range in which Richardson's theory holds. In this sense, eq.~\eqref{eq:alpha_Richardson} can be understood as a relation that matches the values of $g$ and the new universal alignment property in the limited range where the validity of both overlap. 
Further theoretical treatment is still needed to fully understand the alignment property.

Determining the time scaling of the separation between particles in the inertial range is one of the long-standing open problems in turbulence. In particular, it is not yet clear whether the elusiveness of empirically confirming Richardson's eq.~\eqref{eq:Richardson} is due to a failure of the assumption of the theory itself (i.e. a scale dependent diffusivity)~\cite{Sokolov2000, Bitane2012, Thalabard2014, Bourgoin2015} or whether they are due to issues in the measurements. Namely, had the Reynolds numbers in measurements been higher and the duration of measurements been longer, would Richardson's classical predictions be recovered then?
Unlike direct scaling measurements, our work shows that the inertial range behavior of $\av{\theta}$ is clearly observed at the Reynolds numbers and measurement durations readily available with current technological capabilities. In particular, the inertial range behavior is robustly observed for particle pairs with $l_0$ values throughout the whole inertial range (Fig.~\ref{fig:theta_vs_time}a). A central difference between $\av{\theta}$ and the direct scaling method is that while $\theta$ depends on instantaneous regulation of the separation process by the flow (namely on alignment properties of the separation velocity), the scaling exponents integrate it in time and thus can accumulate deviations. Therefore, while our observation that $\av{\theta}$ is constant in the inertial range suggests that the internal regulation of the flow leading to Richardson's law exists in realistic, terrestrial turbulence, whether this leads to converged scaling exponents in a finite Reynolds number flow is still debated~\cite{Kishi2022}.

An important outcome of our work is that it opens the opportunity for verifying the three regimes of turbulent pair dispersion in future measurements. In particular, measuring the evolution of $\av{\theta}$ instead of the separation scaling exponents is advantageous because of three main reasons. First, confirming the super diffusive regime using $\av{\theta}$ amounts to measuring a constant value, which, due to the finite nature of measurements, is much easier than measuring scaling regimes with high exponents. Second, $\theta$ is inherently scale-invariant, so the complications that arise due to the intermittency of pair dispersion and initial separation dependence are avoided. Third, the plateau of $\av{\theta}$ appears early after the ballistic regime (for $t\gtrsim0.5\tau_b$, Fig.~\ref{fig:theta_vs_time}), so it is not necessary to obtain very long particle trajectories. Indeed, the plateau of $\langle \theta \rangle$ is detected for all initial separations inside the inertial range ($l_0<L$). 
Therefore, our framework allows confirming the effects of turbulence on scalar dispersion over a wide range of flows and under realistic conditions - a crucial step both for our understanding of turbulent flows and for modeling dispersion.

To conclude, our geometrical framework opens a new perspective in turbulent dispersion research since its universality allows characterizing pair dispersion in a wide range of turbulent flows. Even under conditions that depart from the ideal ones considered here (i.e., quasi-homogeneous and isotropic), observing at which times and initial separations $\av{\theta}$ remain at a constant value allows to quantitatively assert the range of scales in which the inertial range manifests, a particularly crucial issue for environmental flows~\cite{Celani2005, Pitton2012, Polanco2018}. Therefore, the framework we propose can be used to characterize a large variety of dispersion processes in nature. For example, going back to the question of how fast particles mix in the ocean, measuring $\av{\theta}$ in field experiments will allow determining at which scales isotropic turbulence phenomenology drives dispersion and under which of the three regimes, thus enabling more accurate dispersion estimations of algae blooms, plastic debris and oil spills in the ocean.

\section{Methods} \label{sec:methods}

\subsection{Empirical dataset}

Our work uses two independent empirical datasets to confirm our analysis - the results of a numerical simulation, and the results of a particle tracking experiment.

\subsubsection{Direct numerical simulation}

We used the results of a Direct numerical Simulation of the Navier-Stokes equation with a large scale, statistically stationary and homogeneous forcing. The simulation was performed, and its results are stored and maintained, by the team of the Johns Hopkins Turbulence Database~\cite{Li2008} (JHTDB). The results of the simulation are stored as Eulerian fields, holding data for a time duration of approximately 5 integral timescales over a periodic cubical domain with $1024^3$ nodes; the Taylor microscale Raynolds number is $Re_\lambda\approx 433$. Since its publication, the JHTDB had become a gold standard and an hypothesis testing tool in the turbulence community.

To study pair dispersion using the results of the JHTDB simulation we calculated trajectories of flow tracers by time integration of the velocity field. For that, we used JHTDB's \verb#get_position# function, following the results of~\cite{Yu2012}. Overall, we integrated approximately $12\,500$ trajectories, each over a time span of three integral timescales. Trajectories were integrated in 63 groups starting from random initial positions and times in the computational domain. By pairing initially close trajectories with one another we obtained a dataset that holds $\sim 6\times10^5$ pairs of trajectories with initial separation inside the inertial range (namely, with $l_0 \lesssim 200\eta$), and approximately $10^6$ trajectory pairs with initial separations larger than that. Fig.~\ref{fig:def} shows a 3D rendering of one such group of trajectories.

In addition to the Lagrangian trajectories, we also downloaded Eulerian velocity samples over a regularly spaced grid. This dataset was used to calculate the first-order absolute structure function (Fig.~\ref{fig:3}b). For this purpose we used samples distributed over three orthogonal planes in the computational domain, taken at evenly spaced intervals over time and space, using approximately $6\eta$ of separation in space and one half integral time scale in time.

\subsubsection{3D particle tracking experiment}

We used an experimental dataset obtained from a 3D Particle Tracking Velocimetry (3D-PTV) experiment. The turbulent flow was generated inside an octagonal cylindrical tank that is constructed of an aluminum frame fitted with eight transparent glass windows (Fig.~\ref{fig:3}a). The tank was filled with approximately 17.5 liters of de-ionized filtered water. Turbulent flow was generated in the tank using eight 45~W DC servo-motors, each connected to a vertical shaft that is fitted with a pair of propellers. Each motor was operated individually using a random actuator that alters the direction of rotation at random time intervals taken from a Poisson distribution with an average of 0.1 s, in a setup similar to those used in Refs.~\cite{Berg2006} and~\cite{Luethi2007}.

The flow was measured using 3D Particle Tracking Velocimetry (3D-PTV)~\cite{Dracos1996}. The water was seeded with 70$\mu m$ fluorescent tracer particles of density $\rho=1030 \, \mathrm{kg \, m^{-3}}$. The particles were then illuminated by an expanded laser beam (537 nm), that was masked to illuminate a prismatic rectangular volume at the center of the tank. Four high-speed digital cameras (Microtron, 1280$\times$1024 pixels) were synchronized and operated at 500~Hz to record images of the tracer particles as they were carried by the flow. The cameras were calibrated using the self-calibration method~\cite{Mass1993} using a calibration target on which 438 points were marked at known locations over three parallel planes. The setup yielded a static calibration uncertainty of approximately 50 $\mu m$ based on the root mean squared calibration error. In addition, the cameras were fitted with high pass optical filters to better visualize the fluorescent particles.

The tracer images were analysed following the 3D-PTV methodology by using our open-source software MyPTV~\cite{Shnapp2022} that is freely available online. The trajectories were then smoothed using a moving third-order polynomial spline with a window size of 11 samples, and 5 samples at the edge of each trajectory were discarded. We analysed 20 seconds of data in total, which is on the order of 100 integral timescales of the flow. Our measurement spanned a measurement volume of 70 mm $\times$ 70 mm $\times$ 40 mm, where each of the 10,000 frames in our post-processed dataset contained approximately $\sim \mathcal{O}(10^3)$ particles. The root mean square of the velocity fluctuations was $u'=58 \, \mathrm{mm \, s^{-1}}$, the integral length scale was $L=20.5 \, \mathrm{mm}$, the dissipation rate was $\epsilon = \frac{1}{2} \frac{u'^3}{L} = 4950 \, \mathrm{mm}^2 \, \mathrm{s}^{-3}$; correspondingly, $\eta = 0.12 \, \mathrm{mm}$, and the Taylor microscale and Reynolds number are $\lambda=\sqrt{15\frac{\nu}{\epsilon}}u'=3.1\,\mathrm{mm}$, $\mathrm{Re}_\lambda = \frac{u'\,\lambda}{\nu}\approx188$ (see supplementary materials for details on these calculations). 
A 3D rendered animation of trajectories from our experiment is shown in Supplementary Video 2. 
The full data set of approximately $\sim \mathcal{O}(6 \times 10^5)$ trajectories can be downloaded in Ref.~\cite{Shnapp2022a}.

%As expected, when plotted as a function of time, $A$ plateaus in the range $\tau_b<t<T$ (where $\tau_b\equiv (l_0^{2}/\epsilon)^{1/3}$ is the so-called Batchelor timescale~\cite{Batchelor1952} and $T$ is the integral time scale of the flow) around an average value of 0.89 for both data sets, with fluctuations only due to statistical noise (Fig. S3). 

\subsection{Calculation of the coefficient $b$} \label{sec:calculating_b}

In equation~\eqref{eq:A}, we calculated the first moment of particle separations, $l$, using theoretical predictions of its PDF and its second moment. In particular, two shapes of the PDF of $l$ were predicted assuming diffusive separation from a point source~\cite{Salazar2009}. The first was obtained by Richardson~\cite{Richardson1926}
\begin{equation}
q_R(l)=\frac{a_{R,1}}{(\pi \langle l^2\rangle)^{3/2}}\cdot \exp\left[-a_{R,2}\left(\frac{l^2}{\langle l^2\rangle} \right)^{1/3}\right] \,\, ,
\label{eq:Richardson_PDF}
\end{equation}
where $a_{R,1}=\frac{429}{70}\sqrt{\frac{143}{2}}$
and $a_{R,2}=(\frac{1287}{8})^{1/3}$, and the second was obtained by Batchelor~\cite{Batchelor1952}
\begin{equation}
q_B=\left(\frac{3}{2 \pi \langle l^2\rangle} \right)^{3/2} \, \exp \left(-\frac{3}{2}\frac{l^2}{\langle l^2\rangle}\right)  \,\, .
\label{eq:Batchelor_PDF}
\end{equation}
In both cases the second moment $\langle l^2 \rangle$ appears explicitly in the expressions, which means that the n$^{\mathrm{th}}$ moment can be expressed as $\langle l^n \rangle \propto \langle l^2 \rangle^{\frac{n}{2}}$. In particular, for both PDFs, the n$^{\mathrm{th}}$ moment is readily calculated by solving 
\begin{equation}
\langle l^n \rangle = \int_0^{\infty} l^n \, q(l) \, 4\pi \, l^2 \,dl
\end{equation}
using the following formula~\cite{Gradshteyn2007}
\begin{equation}
\int_0^{\infty} x^n\, \exp(-a\,x^k) \, dx = \frac{1}{k}a^{-\frac{(n+1)}{k}} \Gamma\left(\frac{n+1}{k}\right)
\end{equation}
where $\Gamma(x)$ is the Gamma function. Thus, for the Richardson PDF, $q_R(l)$, we obtain $b = \frac{6 a_{R,1}}{\sqrt{\pi} a_{R,2}^{6}}\Gamma(6) \approx 0.813$, whereas for the Batchelor PDF, $q_{B}(l)$, we obtain $b = \sqrt{ \frac{27}{2\pi} } \, (\frac{2}{3})^{2} \, \Gamma(2)\approx 0.921$. These calculations result in the following range of $b = 0.867 \pm 0.054$. Notably, from the empirical point of view, the measured distribution of $l$ may have a different shape, so $q_R$ and $q_B$ are considered as two limit cases~\cite{Ott2000, Ouellette2006} reflected in the uncertainty range of $b$; in accordance, direct estimation of $b$ using our empirical datasets gives $b\approx0.84$, in agreement with the calculations.

\subsection{The mixed Eulerian-Lagrangian structure function} \label{sec:Eu-Lag_SF}

We here derive the mixed-Eulerian-Lagrangian first order absolute structure function. For that, we consider an ensemble of pairs of Lagrangian particles that are initially separated by a distance $l_0$. The particles are free to move, so the distance between each of the pairs in the ensemble changes as a function of time. Thus, the first order structure function is obtained from the ensemble average as
\begin{equation}
\av{\abs{\Delta \v{v}}} = \iint \abs{\Delta \v{v}} \, \mathcal{P}(\abs{\Delta \v{v}}, l) \, dl \, d\abs{\Delta \v{v}}
%\label{eq:dv_average}
\end{equation}
where $\mathcal{P}(\abs{\Delta \v{v}}, l)$ is the joint PDF for the absolute value of the pairwise relative velocity and the distance (see Ref.~\cite{Monin1972}, Sec. 24.2). Using Bayes' theorem~\cite{VanKampen2007}, the join PDF is written using the marginal and the conditional PDF of the distances and the relative velocity respectively
$\mathcal{P}(\abs{\Delta \v{v}}, \, l) = \mathcal{P}(\abs{\Delta \v{v}} \, \big\lvert \, l) \cdot q(l)\, 4\pi \, l^2 $,
where as in Sec.~\ref{sec:calculating_b}, $q(l)$ is the PDF for the distance $l$. We first solve the integral over the relative velocities and obtain
\begin{equation}
\int \left[ \int \abs{\Delta \v{v}} \, \mathcal{P}(\abs{\Delta \v{v}} \, \big\lvert \, l) \, d\abs{\Delta \v{v}} \right] \, q(l) \, 4\pi \, l^2 \, dl = 
\int C_1 \, \epsilon^{1/3} \, l^{1/3} \, q(l) \, 4\pi \, l^2 \, dl
\label{eq:average_Eu_Lag}
\end{equation}
where we have used the fact the the average of $\abs{\Delta \v{v}}$ at a given scale $l$ is the purely Eulerian first order structure function, while employing Kolmogorov's universal similarity theory~\cite{Kolmogorov1941} for $l$ in the inertial range; this is justified because the initial position of each pair is chosen randomly over space and time, and because of the $l_0$ independence in the inertial range. Following that, we solve the integral over the particle distances while employing both Richardson's and Batchelor's PDFs, $q_R(l)$ and $q_B(l)$. In both cases, we obtain the solution in the form
\begin{equation}
\av{\abs{\Delta \v{v}}} = c \, C_1 \big(\epsilon \, \sqrt{\av{l^2}} \,\big)^{1/3}
\label{eq:mixed_Eu_Lag_structurefunction}
\end{equation}
where $c$ is a constant that depends on the shape of $q(l)$ that was chosen, being equal to 0.952 for the Batchelor PDF, and 0.885 for the Richardson PDF. Thus, we will use the value $c=0.918\pm0.034$ where the uncertainty is taken to make sure we are covering the two possible shapes of the PDF.

\subsection{A Taylor series expansion for $\av{\alpha}$ in the inertial range} \label{sec:alpha_expansion}

For ease of notation, we denote $X \equiv \abs{\Delta \v v}$ and $Y \equiv \frac{d\,l}{dt}$. To obtain eq.~\eqref{eq:A}, we write $\alpha(X,\,Y)=\frac{1}{X} Y$, and Taylor expand it around the averages $\av{X}$ and $\av{Y}$. After averaging the result, we obtain~\cite{Benaroya2005}
\begin{equation}
\begin{split}
\av{\alpha} &= \alpha_0 + \alpha_1 + \alpha_2 + \ldots \\
\alpha_0 &= \alpha\big\lvert_{\av{X},\, \av{Y}} = \frac{1}{\av{X}} \av{Y}\\
\alpha_1 &= \frac{\partial \alpha}{\partial X}\Big\lvert_{\av{X},\, \av{Y}} \, \big\langle X - \av{X}\big\rangle + \frac{\partial \alpha}{\partial Y}\Big\lvert_{\av{X},\, \av{Y}} \big\langle Y - \av{Y} \big\rangle = 0\\
\vdots& \\
\alpha_n &= \big\langle {\frac{1}{n!} \left[ \sum\limits_{k=0}^n \binom{n}{k} \, \frac{\partial^{n-k}}{\partial X^{n-k}} \, \frac{\partial^{k} \alpha}{\partial Y^{k}} \Big\lvert_{\av{X}, \, \av{Y}} \,  (X-\av{X})^{n-k} (Y-\av{Y})^{k} \right]\big\rangle}
\end{split} 
\end{equation}
Let us note that the values of $\alpha$ and its derivatives estimated at $\av{X}$ and $\av{Y}$ are constants that multiply mixed central moments of $X$ and $Y$.

\bmhead{Data availability}

The experimental dataset generated and analyzed during the current study is available in "Zenodo" with the identifier \href{https://doi.org/10.5281/zenodo.6802679}{https://doi.org/10.5281/zenodo.6802679}. Numerical data from the DNS dataset used in this study can be downloaded from the Johns Hopkins Turbulence Database at \href{http://turbulence.pha.jhu.edu/}{http://turbulence.pha.jhu.edu/}. Other data used to support the finding of this study are available from the authors upon reasonable request.

% =========================================
% ============ Bibliography: ==============

%% BioMed_Central_Bib_Style_v1.01

%\bibliography{bib.bib}

\begin{thebibliography}{41}
	% BibTex style file: bmc-mathphys.bst (version 2.1), 2014-07-24
	\ifx \bisbn   \undefined \def \bisbn  #1{ISBN #1}\fi
	\ifx \binits  \undefined \def \binits#1{#1}\fi
	\ifx \bauthor  \undefined \def \bauthor#1{#1}\fi
	\ifx \batitle  \undefined \def \batitle#1{#1}\fi
	\ifx \bjtitle  \undefined \def \bjtitle#1{#1}\fi
	\ifx \bvolume  \undefined \def \bvolume#1{\textbf{#1}}\fi
	\ifx \byear  \undefined \def \byear#1{#1}\fi
	\ifx \bissue  \undefined \def \bissue#1{#1}\fi
	\ifx \bfpage  \undefined \def \bfpage#1{#1}\fi
	\ifx \blpage  \undefined \def \blpage #1{#1}\fi
	\ifx \burl  \undefined \def \burl#1{\textsf{#1}}\fi
	\ifx \doiurl  \undefined \def \doiurl#1{\url{https://doi.org/#1}}\fi
	\ifx \betal  \undefined \def \betal{\textit{et al.}}\fi
	\ifx \binstitute  \undefined \def \binstitute#1{#1}\fi
	\ifx \binstitutionaled  \undefined \def \binstitutionaled#1{#1}\fi
	\ifx \bctitle  \undefined \def \bctitle#1{#1}\fi
	\ifx \beditor  \undefined \def \beditor#1{#1}\fi
	\ifx \bpublisher  \undefined \def \bpublisher#1{#1}\fi
	\ifx \bbtitle  \undefined \def \bbtitle#1{#1}\fi
	\ifx \bedition  \undefined \def \bedition#1{#1}\fi
	\ifx \bseriesno  \undefined \def \bseriesno#1{#1}\fi
	\ifx \blocation  \undefined \def \blocation#1{#1}\fi
	\ifx \bsertitle  \undefined \def \bsertitle#1{#1}\fi
	\ifx \bsnm \undefined \def \bsnm#1{#1}\fi
	\ifx \bsuffix \undefined \def \bsuffix#1{#1}\fi
	\ifx \bparticle \undefined \def \bparticle#1{#1}\fi
	\ifx \barticle \undefined \def \barticle#1{#1}\fi
	\bibcommenthead
	\ifx \bconfdate \undefined \def \bconfdate #1{#1}\fi
	\ifx \botherref \undefined \def \botherref #1{#1}\fi
	\ifx \url \undefined \def \url#1{\textsf{#1}}\fi
	\ifx \bchapter \undefined \def \bchapter#1{#1}\fi
	\ifx \bbook \undefined \def \bbook#1{#1}\fi
	\ifx \bcomment \undefined \def \bcomment#1{#1}\fi
	\ifx \oauthor \undefined \def \oauthor#1{#1}\fi
	\ifx \citeauthoryear \undefined \def \citeauthoryear#1{#1}\fi
	\ifx \endbibitem  \undefined \def \endbibitem {}\fi
	\ifx \bconflocation  \undefined \def \bconflocation#1{#1}\fi
	\ifx \arxivurl  \undefined \def \arxivurl#1{\textsf{#1}}\fi
	\csname PreBibitemsHook\endcsname
	
	%%% 1
	\bibitem{Batchelor1952}
	\begin{barticle}
		\bauthor{\bsnm{Batchelor}, \binits{G.K.}}:
		\batitle{Diffusion in a field of homogeneous turbulence: {II}. {T}he relative
			motion of particles}.
		\bjtitle{Mathematical Proceedings of the Cambridge Philosophical Society}
		\bvolume{48}(\bissue{2}),
		\bfpage{345}--\blpage{362}
		(\byear{1952}).
		\doiurl{10.1017/S0305004100027687}
	\end{barticle}
	\endbibitem
	
	%%% 2
	\bibitem{Edouard1996}
	\begin{barticle}
		\bauthor{\bsnm{Edouard}, \binits{S.}},
		\bauthor{\bsnm{Legras}, \binits{B.}},
		\bauthor{\bsnm{Lefevre}, \binits{F.}},
		\bauthor{\bsnm{Eymard}, \binits{R.}}:
		\batitle{The effect of small-scale inhomogeneities on ozone depletion in the
			{Arctic}}.
		\bjtitle{Nature}
		\bvolume{384}(\bissue{6608}),
		\bfpage{444}--\blpage{447}
		(\byear{1996}).
		\doiurl{10.1038/384444a0}
	\end{barticle}
	\endbibitem
	
	%%% 3
	\bibitem{Griffa2013}
	\begin{barticle}
		\bauthor{\bsnm{Griffa}, \binits{A.}},
		\bauthor{\bsnm{Haza}, \binits{A.}},
		\bauthor{\bsnm{Oezgoekmen}, \binits{T.M.}},
		\bauthor{\bsnm{Molcard}, \binits{A.}},
		\bauthor{\bsnm{Taillandier}, \binits{V.}},
		\bauthor{\bsnm{Schroeder}, \binits{K.}},
		\bauthor{\bsnm{Chang}, \binits{Y.}},
		\bauthor{\bsnm{Poulain}, \binits{P.-M.}}:
		\batitle{Investigating transport pathways in the ocean}.
		\bjtitle{Deep Sea Research Part II: Topical Studies in Oceanography}
		\bvolume{85},
		\bfpage{81}--\blpage{95}
		(\byear{2013}).
		\doiurl{10.1016/j.dsr2.2012.07.031}
	\end{barticle}
	\endbibitem
	
	%%% 4
	\bibitem{Monin1972}
	\begin{bbook}
		\bauthor{\bsnm{Monin}, \binits{A.S.}},
		\bauthor{\bsnm{Yaglom}, \binits{A.M.}}:
		\bbtitle{{Statistical Fluid Mechanics}}.
		\bpublisher{Dover Publications inc.},
		\blocation{Mineola, N.Y.}
		(\byear{1972}).
		\doiurl{10.1119/1.10870}
	\end{bbook}
	\endbibitem
	
	%%% 5
	\bibitem{Sawford2001}
	\begin{barticle}
		\bauthor{\bsnm{Sawford}, \binits{B.}}:
		\batitle{Turbulent relative dispersion}.
		\bjtitle{Annual Review of Fluid Mechanics}
		\bvolume{33}(\bissue{1}),
		\bfpage{289}--\blpage{317}
		(\byear{2001})
		{\href{https://arxiv.org/abs/https://doi.org/10.1146/annurev.fluid.33.1.289}{{https://doi.org/10.1146/annurev.fluid.33.1.289}}}.
		\doiurl{10.1146/annurev.fluid.33.1.289}
	\end{barticle}
	\endbibitem
	
	%%% 6
	\bibitem{Salazar2009}
	\begin{barticle}
		\bauthor{\bsnm{Salazar}, \binits{J.P.L.C.}},
		\bauthor{\bsnm{Collins}, \binits{L.R.}}:
		\batitle{Two-particle dispersion in isotropic turbulent flows}.
		\bjtitle{Annual review of fluid mechanics}
		\bvolume{41}(\bissue{1}),
		\bfpage{405}--\blpage{432}
		(\byear{2009}).
		\doiurl{10.1146/annurev.fluid.40.111406.102224}
	\end{barticle}
	\endbibitem
	
	%%% 7
	\bibitem{Taylor1921}
	\begin{barticle}
		\bauthor{\bsnm{Taylor}, \binits{G.I.}}:
		\batitle{Diffusion by continuous movements}.
		\bjtitle{Proceedings of the London Mathematical Society}
		(\byear{1921}).
		\doiurl{10.1112/plms/s2-20.1.196}
	\end{barticle}
	\endbibitem
	
	%%% 8
	\bibitem{Richardson1926}
	\begin{barticle}
		\bauthor{\bsnm{Richardson}, \binits{L.F.}}:
		\batitle{Atmospheric diffusion shown on a distance-neighbour graph}.
		\bjtitle{Proceedings of the Royal Society of London. Series A, Containing
			Papers of a Mathematical and Physical Character}
		\bvolume{110}(\bissue{756}),
		\bfpage{709}--\blpage{737}
		(\byear{1926}).
		\doiurl{10.1098/rspa.1926.0043}
	\end{barticle}
	\endbibitem
	
	%%% 9
	\bibitem{Yeung1994}
	\begin{barticle}
		\bauthor{\bsnm{Yeung}, \binits{P.K.}}:
		\batitle{Direct numerical simulation of two-particle relative diffusion in
			isotropic turbulence}.
		\bjtitle{Physics of Fluids}
		\bvolume{6}(\bissue{10}),
		\bfpage{3416}--\blpage{3428}
		(\byear{1994}).
		\doiurl{10.1063/1.868399}
	\end{barticle}
	\endbibitem
	
	%%% 10
	\bibitem{Ott2000}
	\begin{barticle}
		\bauthor{\bsnm{Ott}, \binits{S.}},
		\bauthor{\bsnm{Mann}, \binits{J.}}:
		\batitle{An experimental investigation of the relative diffusion of particle pairs in three-dimensional turbulent flow}.
		\bjtitle{Journal of Fluid Mechanics}
		\bvolume{422},
		\bfpage{207}--\blpage{223}
		(\byear{2000}).
		\doiurl{10.1017/S0022112000001658}
	\end{barticle}
	\endbibitem
	
	%%% 11
	\bibitem{Biferale2005}
	\begin{barticle}
		\bauthor{\bsnm{Biferale}, \binits{L.}},
		\bauthor{\bsnm{Boffetta}, \binits{G.}},
		\bauthor{\bsnm{Celani}, \binits{A.}},
		\bauthor{\bsnm{Devenish}, \binits{B.J.}},
		\bauthor{\bsnm{Lanotte}, \binits{A.}},
		\bauthor{\bsnm{Toschi}, \binits{F.}}:
		\batitle{Lagrangian statistics of particle pairs in homogeneous isotropic
			turbulence}.
		\bjtitle{Physics of Fluids}
		\bvolume{17}(\bissue{11}),
		\bfpage{115101}
		(\byear{2005}).
		\doiurl{10.1063/1.2130742}
	\end{barticle}
	\endbibitem
	
	%%% 12
	\bibitem{Bourgoin2006}
	\begin{barticle}
		\bauthor{\bsnm{Bourgoin}, \binits{M.}},
		\bauthor{\bsnm{Ouellette}, \binits{N.T.}},
		\bauthor{\bsnm{Xu}, \binits{H.}},
		\bauthor{\bsnm{Berg}, \binits{J.}},
		\bauthor{\bsnm{Bodenschatz}, \binits{E.}}:
		\batitle{The role of pair dispersion in turbulent flow}.
		\bjtitle{Science}
		\bvolume{311},
		\bfpage{835}--\blpage{838}
		(\byear{2006}).
		\doiurl{10.1126/science.1121726}
	\end{barticle}
	\endbibitem
	
	%%% 13
	\bibitem{Ouellette2006}
	\begin{botherref}
		\oauthor{\bsnm{Ouellette}, \binits{N.T.}},
		\oauthor{\bsnm{Xu}, \binits{H.}},
		\oauthor{\bsnm{Bourgoin}, \binits{M.}},
		\oauthor{\bsnm{Bodenschatz}, \binits{E.}}:
		An experimental study of turbulent relative dispersion models.
		New Journal of Physics
		\textbf{8}
		(2006).
		\doiurl{10.1088/1367-2630/8/6/109}
	\end{botherref}
	\endbibitem
	
	%%% 14
	\bibitem{Buaria2015}
	\begin{barticle}
		\bauthor{\bsnm{Buaria}, \binits{D.}},
		\bauthor{\bsnm{Sawford}, \binits{B.L.}},
		\bauthor{\bsnm{Yeung}, \binits{P.K.}}:
		\batitle{Characteristics of backward and forward two-particle relative
			dispersion in turbulence at different {Reynolds} numbers}.
		\bjtitle{Physics of Fluids}
		\bvolume{27}(\bissue{10}),
		\bfpage{105101}
		(\byear{2015}).
		\doiurl{10.1063/1.4931602}
	\end{barticle}
	\endbibitem
	
	%%% 15
	\bibitem{Shnapp2018}
	\begin{barticle}
		\bauthor{\bsnm{Shnapp}, \binits{R.}},
		\bauthor{\bsnm{Liberzon}, \binits{A.}}:
		\batitle{Generalization of turbulent pair dispersion to large initial
			separations}.
		\bjtitle{Phys. Rev. Lett.}
		\bvolume{120},
		\bfpage{244502}
		(\byear{2018}).
		\doiurl{10.1103/PhysRevLett.120.244502}
	\end{barticle}
	\endbibitem
	
	%%% 16
	\bibitem{Tan2022}
	\begin{barticle}
		\bauthor{\bsnm{Tan}, \binits{S.}},
		\bauthor{\bsnm{Ni}, \binits{R.}}:
		\batitle{Universality and intermittency of pair dispersion in turbulence}.
		\bjtitle{Phys. Rev. Lett.}
		\bvolume{128},
		\bfpage{114502}
		(\byear{2022}).
		\doiurl{10.1103/PhysRevLett.128.114502}
	\end{barticle}
	\endbibitem
	
	%%% 17
	\bibitem{Elsinga2022}
	\begin{barticle}
		\bauthor{\bsnm{Elsinga}, \binits{G.E.}},
		\bauthor{\bsnm{Ishihara}, \binits{T.}},
		\bauthor{\bsnm{Hunt}, \binits{J.C.R.}}:
		\batitle{Non-local dispersion and the reassessment of {Richardson's} t3-scaling
			law}.
		\bjtitle{Journal of Fluid Mechanics}
		\bvolume{932},
		\bfpage{17}
		(\byear{2022}).
		\doiurl{10.1017/jfm.2021.989}
	\end{barticle}
	\endbibitem
	
	%%% 18
	\bibitem{Boffetta2002}
	\begin{botherref}
		\oauthor{\bsnm{Boffetta}, \binits{G.}},
		\oauthor{\bsnm{Sokolov}, \binits{I.M.}}:
		Relative dispersion in fully developed turbulence: the {Richardson's} law and
		intermittency corrections.
		Physical Review Letters
		\textbf{88}
		(2002).
		\doiurl{10.1103/PhysRevLett.88.094501}
	\end{botherref}
	\endbibitem
	
	%%% 19
	\bibitem{Yeung2004}
	\begin{barticle}
		\bauthor{\bsnm{Yeung}, \binits{P.K.}},
		\bauthor{\bsnm{Borgas}, \binits{M.S.}}:
		\batitle{Relative dispersion in isotropic turbulence. {Part 1. Direct}
			numerical simulations and {Reynolds}-number dependence}.
		\bjtitle{Journal of Fluid Mechanics}
		\bvolume{503},
		\bfpage{93}--\blpage{124}
		(\byear{2004}).
		\doiurl{10.1017/S0022112003007584}
	\end{barticle}
	\endbibitem
	
	%%% 20
	\bibitem{Scatamacchia2012}
	\begin{barticle}
		\bauthor{\bsnm{Scatamacchia}, \binits{R.}},
		\bauthor{\bsnm{Biferale}, \binits{L.}},
		\bauthor{\bsnm{Toschi}, \binits{F.}}:
		\batitle{Extreme events in the dispersions of two neighboring particles under
			the influence of fluid turbulence}.
		\bjtitle{Phys. Rev. Lett.}
		\bvolume{109},
		\bfpage{144501}
		(\byear{2012}).
		\doiurl{10.1103/PhysRevLett.109.144501}
	\end{barticle}
	\endbibitem
	
	%%% 21
	\bibitem{Berg2006}
	\begin{barticle}
		\bauthor{\bsnm{Berg}, \binits{J.}},
		\bauthor{\bsnm{L\"uthi}, \binits{B.}},
		\bauthor{\bsnm{Mann}, \binits{J.}},
		\bauthor{\bsnm{Ott}, \binits{S.}}:
		\batitle{Backwards and forwards relative dispersion in turbulent flow: An
			experimental investigation}.
		\bjtitle{Phys. Rev. E}
		\bvolume{74},
		\bfpage{016304}
		(\byear{2006}).
		\doiurl{10.1103/PhysRevE.74.016304}
	\end{barticle}
	\endbibitem
	
	%%% 22
	\bibitem{Kishi2022}
	\begin{barticle}
		\bauthor{\bsnm{Kishi}, \binits{T.}},
		\bauthor{\bsnm{Matsumoto}, \binits{T.}},
		\bauthor{\bsnm{Toh}, \binits{S.}}:
		\batitle{Two-time {Lagrangian} velocity correlation function for particle pairs
			in two-dimensional inverse energy-cascade turbulence}.
		\bjtitle{Physical Review Fluids}
		\bvolume{7}(\bissue{6}),
		\bfpage{064604}
		(\byear{2022}).
		\doiurl{10.1103/physrevfluids.7.064604}
	\end{barticle}
	\endbibitem
	
	%%% 23
	\bibitem{Tsinober2009}
	\begin{bbook}
		\bauthor{\bsnm{Tsinober}, \binits{A.}}:
		\bbtitle{An Informal Conceptual Introduction to Turbulence}.
		\bpublisher{Springer}
		(\byear{2009}).
		\doiurl{10.1007/978-90-481-3174-7_6}
	\end{bbook}
	\endbibitem
	
	%%% 24
	\bibitem{Kolmogorov1941}
	\begin{barticle}
		\bauthor{\bsnm{Kolmogorov}, \binits{A.N.}}:
		\batitle{The local structure of turbulence in incompressible viscous fluid for
			very large {Reynolds} numbers}.
		\bjtitle{Cr Acad. Sci. URSS}
		\bvolume{30},
		\bfpage{301}--\blpage{305}
		(\byear{1941})
	\end{barticle}
	\endbibitem
	
	%%% 25
	\bibitem{Li2008}
	\begin{botherref}
		\oauthor{\bsnm{Li}, \binits{Y.}},
		\oauthor{\bsnm{Perlman}, \binits{E.}},
		\oauthor{\bsnm{Wan}, \binits{M.}},
		\oauthor{\bsnm{Yang}, \binits{Y.}},
		\oauthor{\bsnm{Meneveau}, \binits{C.}},
		\oauthor{\bsnm{Burns}, \binits{R.}},
		\oauthor{\bsnm{Chen}, \binits{S.}},
		\oauthor{\bsnm{Szalay}, \binits{A.}},
		\oauthor{\bsnm{Eyink}, \binits{G.}}:
		A public turbulence database cluster and applications to study {Lagrangian}
		evolution of velocity increments in turbulence.
		Journal of Turbulence
		(9),
		31
		(2008).
		\doiurl{10.1080/14685240802376389}
	\end{botherref}
	\endbibitem
	
	%%% 26
	\bibitem{Yu2012}
	\begin{botherref}
		\oauthor{\bsnm{Yu}, \binits{H.}},
		\oauthor{\bsnm{Kanov}, \binits{K.}},
		\oauthor{\bsnm{Perlman}, \binits{E.}},
		\oauthor{\bsnm{Graham}, \binits{J.}},
		\oauthor{\bsnm{Frederix}, \binits{E.}},
		\oauthor{\bsnm{Burns}, \binits{R.}},
		\oauthor{\bsnm{Szalay}, \binits{A.}},
		\oauthor{\bsnm{Eyink}, \binits{G.}},
		\oauthor{\bsnm{Meneveau}, \binits{C.}}:
		Studying {Lagrangian} dynamics of turbulence using on-demand fluid particle
		tracking in a public turbulence database.
		Journal of Turbulence
		(13),
		12
		(2012)
	\end{botherref}
	\endbibitem
	
	%%% 27
	\bibitem{Dracos1996}
	\begin{bbook}
		\bauthor{\bsnm{Dracos}, \binits{T.}}:
		\bbtitle{Three-dimensional velocity and vorticity measuring and image analysis
			technique: Lecture notes from the short course held in Zurich, Switzerland}.
		\bpublisher{Springer}
		(\byear{1996}).
		\burl{https://www.springer.com/gp/book/9780792342564}
	\end{bbook}
	\endbibitem
	
	%%% 28
	\bibitem{Shnapp2022}
	\begin{barticle}
		\bauthor{\bsnm{Shnapp}, \binits{R.}}:
		\batitle{{MyPTV}: A {Python} package for {3D} particle tracking}.
		\bjtitle{Journal of Open Source Software}
		\bvolume{7}(\bissue{75}),
		\bfpage{4398}
		(\byear{2022}).
		\doiurl{10.21105/joss.04398}
	\end{barticle}
	\endbibitem
	
	%%% 29
	\bibitem{Shnapp2022a}
	\begin{botherref}
		\oauthor{\bsnm{Shnapp}, \binits{R.}},
		\oauthor{\bsnm{Brizzolara}, \binits{S.}},
		\oauthor{\bsnm{Neamtu~Halic}, \binits{M.}},
		\oauthor{\bsnm{Gambino}, \binits{A.}},
		\oauthor{\bsnm{Holzner}, \binits{M.}}:
		Lagrangian particles in turbulence: An experimental data set.
		Zenodo
		(2022).
		\doiurl{10.5281/zenodo.6802680}
	\end{botherref}
	\endbibitem
	
	%%% 30
	\bibitem{Bitane2012}
	\begin{barticle}
		\bauthor{\bsnm{Bitane}, \binits{R.}},
		\bauthor{\bsnm{Homann}, \binits{H.}},
		\bauthor{\bsnm{Bec}, \binits{J.}}:
		\batitle{Time scales of turbulent relative dispersion}.
		\bjtitle{Phys. Rev. E}
		\bvolume{86},
		\bfpage{045302}
		(\byear{2012}).
		\doiurl{10.1103/PhysRevE.86.045302}
	\end{barticle}
	\endbibitem
	
	%%% 31
	\bibitem{Sokolov2000}
	\begin{barticle}
		\bauthor{\bsnm{Sokolov}, \binits{I.M.}},
		\bauthor{\bsnm{Klafter}, \binits{J.}},
		\bauthor{\bsnm{Blumen}, \binits{A.}}:
		\batitle{Ballistic versus diffusive pair dispersion in the {Richardson} regime}.
		\bjtitle{Phys. Rev. E}
		\bvolume{61},
		\bfpage{2717}--\blpage{2722}
		(\byear{2000}).
		\doiurl{10.1103/PhysRevE.61.2717}
	\end{barticle}
	\endbibitem
	
	%%% 32
	\bibitem{Thalabard2014}
	\begin{barticle}
		\bauthor{\bsnm{Thalabard}, \binits{S.}},
		\bauthor{\bsnm{Krstulovic}, \binits{G.}},
		\bauthor{\bsnm{Bec}, \binits{J.}}:
		\batitle{Turbulent pair dispersion as a continuous-time random walk}.
		\bjtitle{Journal of Fluid Mechanics}
		\bvolume{755},
		\bfpage{4}
		(\byear{2014}).
		\doiurl{10.1017/jfm.2014.445}
	\end{barticle}
	\endbibitem
	
	%%% 33
	\bibitem{Bourgoin2015}
	\begin{barticle}
		\bauthor{\bsnm{Bourgoin}, \binits{M.}}:
		\batitle{Turbulent pair dispersion as a ballistic cascade phenomenology}.
		\bjtitle{Journal of Fluid Mechanics}
		\bvolume{772},
		\bfpage{678}--\blpage{704}
		(\byear{2015}).
		\doiurl{10.1017/jfm.2015.206}
	\end{barticle}
	\endbibitem
	
	%%% 34
	\bibitem{Celani2005}
	\begin{barticle}
		\bauthor{\bsnm{Celani}, \binits{A.}},
		\bauthor{\bsnm{Cencini}, \binits{M.}},
		\bauthor{\bsnm{Vergassola}, \binits{M.}},
		\bauthor{\bsnm{Villermaux}, \binits{E.}},
		\bauthor{\bsnm{Vincenzi}, \binits{D.}}:
		\batitle{Shear effects on passive scalar spectra}.
		\bjtitle{Journal of Fluid Mechanics}
		\bvolume{523},
		\bfpage{99}--\blpage{108}
		(\byear{2005}).
		\doiurl{10.1017/S0022112004002332}
	\end{barticle}
	\endbibitem
	
	%%% 35
	\bibitem{Pitton2012}
	\begin{barticle}
		\bauthor{\bsnm{Pitton}, \binits{E.}},
		\bauthor{\bsnm{Marchioli}, \binits{C.}},
		\bauthor{\bsnm{Lavezzo}, \binits{V.}},
		\bauthor{\bsnm{Soldati}, \binits{A.}},
		\bauthor{\bsnm{Toschi}, \binits{F.}}:
		\batitle{Anisotropy in pair dispersion of inertial particles in turbulent
			channel flow}.
		\bjtitle{Physics of Fluids}
		\bvolume{24}(\bissue{7}),
		\bfpage{073305}
		(\byear{2012})
		{\href{https://arxiv.org/abs/https://doi.org/10.1063/1.4737655}{{https://doi.org/10.1063/1.4737655}}}.
		\doiurl{10.1063/1.4737655}
	\end{barticle}
	\endbibitem
	
	%%% 36
	\bibitem{Polanco2018}
	\begin{barticle}
		\bauthor{\bsnm{Polanco}, \binits{J.I.}},
		\bauthor{\bsnm{Vinkovic}, \binits{I.}},
		\bauthor{\bsnm{Stelzenmuller}, \binits{N.}},
		\bauthor{\bsnm{Mordant}, \binits{N.}},
		\bauthor{\bsnm{Bourgoin}, \binits{M.}}:
		\batitle{Relative dispersion of particle pairs in turbulent channel flow}.
		\bjtitle{International Journal of Heat and Fluid Flow}
		\bvolume{71},
		\bfpage{231}--\blpage{245}
		(\byear{2018}).
		\doiurl{10.1016/j.ijheatfluidflow.2018.04.007}
	\end{barticle}
	\endbibitem
	
	%%% 37
	\bibitem{Luethi2007}
	\begin{botherref}
		\oauthor{\bsnm{L{\"u}thi}, \binits{B.}},
		\oauthor{\bsnm{Ott}, \binits{S.}},
		\oauthor{\bsnm{Berg}, \binits{J.}},
		\oauthor{\bsnm{Mann}, \binits{J.}}:
		\batitle{Lagrangian multi-particle statistics}.
		\bjtitle{Journal of Turbulence}
		\bvolume{8},
		\bfpage{N45}
		(\byear{2007}).
		\doiurl{10.1080/14685240701522927}
	\end{botherref}
	\endbibitem
	
	%%% 38
	\bibitem{Mass1993}
	\begin{barticle}
		\bauthor{\bsnm{Mass}, \binits{H.G.}},
		\bauthor{\bsnm{Gruen}, \binits{D.}},
		\bauthor{\bsnm{Papantoniou}, \binits{D.}}:
		\batitle{Particle tracking velocimetry in three-dimensional flows part i:
			Photogrammetric determination of particle coordinates}.
		\bjtitle{Experiments in Fluid}
		\bvolume{15},
		\bfpage{133}--\blpage{146}
		(\byear{1993}).
		\doiurl{10.1007/BF00190953}
	\end{barticle}
	\endbibitem
	
	%%% 39
	\bibitem{Gradshteyn2007}
	\begin{bbook}
		\bauthor{\bsnm{Gradshteyn}, \binits{I.S.}},
		\bauthor{\bsnm{Ryzhik}, \binits{I.M.}}:
		\bbtitle{Table of Integrals, Series, and Products},
		\bedition{7}th edn.
		\bpublisher{Academic Press}
		(\byear{2007}).
		\doiurl{10.1016/C2010-0-64839-5}
	\end{bbook}
	\endbibitem
	
	%%% 40
	\bibitem{VanKampen2007}
	\begin{bbook}
		\bauthor{\bsnm{Van~Kampen}, \binits{N.G.}}:
		\bbtitle{Stochastic Processes in Physics and Chemistry}.
		\bpublisher{Elsevier}
		(\byear{2007}).
		\doiurl{10.1016/B978-0-444-52965-7.X5000-4}
	\end{bbook}
	\endbibitem
	
	%%% 41
	\bibitem{Benaroya2005}
	\begin{bbook}
		\bauthor{\bsnm{Benaroya}, \binits{H.}},
		\bauthor{\bsnm{Han}, \binits{S.M.}},
		\bauthor{\bsnm{Nagurka}, \binits{M.}}:
		\bbtitle{Probability Models in Engineering and Science}
		vol. \bseriesno{192}.
		\bpublisher{CRC press}
		(\byear{2005})
	\end{bbook}
	\endbibitem
	
\end{thebibliography}

% ============ End of bibliography ==============
% ===============================================

\backmatter

%
%\bmhead{Supplementary information}
%
%Two supplementary videos are available for this paper: 
%\begin{enumerate}
%	\item \textit{Supplementary\_video1}, visualizes the angle formed between the relative position and relative velocity of two particles in the DNS dataset;
%	\item \textit{Supplementary\_video2}, visualizes particle trajectories taken from the experimental 3D-PTV measurements.
%\end{enumerate}
%In addition to that, further information on pair dispersion in the experimental dataset and the code for obtaining the data from the JHTDB is shown in a supplementary information file. 
%

\bmhead{Acknowledgments}

We would like to thank Prof. Guido Boffetta and Dr. Simone Boi for helpful comments and discussions. R.S. is a Rothschild fellow. M.H. and S.B. acknowledge financial support from the DFG priority program SPP 1881 Turbulent Superstructures under Grant No. HO5519/1-2.

%
%
%\bmhead{Authors' contributions}
%
%R.S. initiated the study. The theory was developed by R.S., S.B., M.M.N.H., A.G. and M.H. The numerical data was obtained and analyzed by R.S. The experiment was designed and performed, and its results were analyzed by R.S., S.B., M.M.N.H. and A.G. The paper was written by R.S., S.B., M.M.N.H., A.G., and M.H. 

\bmhead{Competing interests}

The authors declare no competing interests.

%\bmhead{Additional Information}
%
%\begin{itemize}
%	\item Supplementary Information is available for this paper.
%	
%	\item Correspondence and requests for materials should be addressed to R.S.
%\end{itemize}
%

\end{document}